\documentstyle[12pt]{article}

\begin{document}

\setcounter{page}{1}

\pagestyle{plain} \vspace{1cm}
\begin{center}
\Large{\bf Phantom-like effects in asymmetric brane embedding with
induced gravity and the Gauss-Bonnet term in the Bulk}\\
\small \vspace{1cm} {\bf Kourosh
Nozari$^{a,b,}$\footnote{knozari@umz.ac.ir}}\quad and \quad
{\bf Tahereh Azizi$^{a,}$\footnote{t.azizi@umz.ac.ir}}\quad  \\
\vspace{0.5cm} {\it $^{a}$Department of Physics, Faculty of Basic
Sciences,\\
University of Mazandaran,\\
P. O. Box 47416-95447, Babolsar, IRAN\\
\vspace{0.5cm} $^{b}$Research Institute for Astronomy and
Astrophysics of Maragha,\\
P. O. Box 55134-441, Maragha, IRAN}
\end{center} \vspace{1cm}
\begin{abstract}
We construct an asymmetric braneworld embedding with induced gravity
on the brane, where stringy effects are taken into account by
incorporation of the Gauss-Bonnet term in the bulk action. We derive
the effective Friedmann equation of the brane and then we
investigate the possible realization of the phantom-like behavior in
this setup. We show that in the absence of the Gauss-Bonnet term in
the bulk action (a pure induced gravity scenario), the phantom-like
behavior in asymmetric case can be realized in smaller redshift than
the corresponding symmetric case. We show also that in the general
case with curvature effect, the phantom-like behavior can be
realized in two subcases: in a symmetric subcase and also in an
asymmetric branch of the solutions. In either cases this
phantom-like behavior happens without introducing any phantom fields
neither on the brane nor in the bulk.
\end{abstract}
\vspace{2cm}
\newpage
\section{Introduction}
An even increasing number of observational data indicate that our
universe is currently in a phase of accelerating expansion [1]. A
simple way to explain the cosmic acceleration is introducing a
cosmological constant, with an equation of state parameter
$\omega_{\Lambda}=-1$, and then investigating the so-called $\Lambda
CDM$ model. This model is favored by recent observations, but it has
some important difficulties such as the fine-tuning and coincidence
problems and also unknown origin of the cosmological constant [2].
To accommodate the cosmic speed up, some other types of unknown
energy components ( the so-called dark energy) with negative
pressure have been proposed [3]. However, understanding the nature
of the dark energy is one of the fundamental problems of modern
theoretical cosmology [4]. There is another approach to realize the
late-time acceleration: modifying geometric part of the
gravitational theory. This proposal can be realized in braneworld
scenarios [5], string inspired scenarios [6], $f(R)$ gravity [7],
and so on.

Since the last decade, there has been a lot of interest in the extra
dimensional theories by modifying the old Kaluza-Klien picture,
where the extra dimensions must be sufficiently compact. These
recent developments are based on the idea that ordinary matter and
gauge fields could be confined to a three dimensional world
(3-brane), while gravity and possibly non-standard matter are free
to propagate in the entire extra dimensional spacetime (the bulk).
The cosmological braneworld solutions exhibit many interesting and
unusual properties such as modifying the gravity at early or late
times. The former issue is presented in the RSII model [8], that a
positive tension brane is embedded in an anti de Sitter (AdS$5$)
bulk and the latter case is associated with the braneworld models
with an induced gravity term in the brane action. The DGP scenario
[5] is a subclass of these models that a tensionless brane is
embedded in an infinite Minkowski bulk. In this model, gravity leaks
off the brane into extra dimension at large scales. Gravity leakage
at late times causes the cosmic speed up due to the weakening of
gravity on the brane, without need to introducing a dark energy
component [9].

Most braneworld scenarios assume a $Z_2$-symmetric brane which is
motivated by a model of M-theory proposed by Horava and Witten [10].
However, some recent papers examine the more general models that are
not directly derived from M-theory, relaxing the mirror symmetry
across the brane (for instance see [11-16]). In the asymmetric case
( without the $Z_2$ symmetry), the parameters in the bulk, such as
the gravitational and cosmological constants, can differ on either
sides of the brane. A distinctive property of the asymmetric brane
model is that they can admit self-accelerating solutions even
without need to introduce an induced gravity term in the brane
action [15,16].

Since the braneworld scenarios are motivated by string theories, it
is natural to include some extra terms such as the Gauss-Bonnet (GB)
term in the five-dimensional field equations. The Gauss-Bonnet
extension of General Relativity (GR) has been motivated from a
string theoretical point of view as a version of higher-dimensional
gravity, since this sort of modification also appears in low energy
effective actions in this context [17]. The GB term leads to
second-order gravitational field equations linear in the second
derivatives in the bulk metric which is ghost free, the property of
curvature invariant of the Gauss-Bonnet term [18]. Inclusion of the
Gauss-Bonnet term in the bulk action results in a variety of novel
phenomena which certainly affects the cosmological dynamics of these
generalized braneworld setup [19].

Recently, well known astronomical observations with WMAP$5$ have
indicated that the equation of state parameter of dark energy can be
less than $-1$ and even can have a transient behavior [20]. A simple
way to explain this phenomenon is to consider a non-canonical
phantom dark energy [21] that introduces new theoretical facilities
and challenges in this field. Phantom fields are a sort of scalar
fields with negative sign for the kinetic energy term. Indeed,
phantom fields suffer from instabilities due to violation of the
null energy condition, and a phantom universe eventually ends up
with a Big Rip singularity [22]. Thus it follows immediately that
there must be some alternative approaches to realize a phantom-like
behavior without introducing any phantom field in the model. With
phantom-like behavior, we mean the growth of the effective dark
energy density with cosmic time and in the same time, the effective
equation of state parameter should stay always less than $-1$. In
this regard, it has been shown that the normal,
non-self-accelerating branch of the DGP scenario has the potential
to explain the phantom-like behavior without introducing any phantom
fields on the brane [23]. This type of the analysis then has been
extended by several authors [24]. The phantom-like behavior of
$4$-dimensional $f(R)$ gravity is studied in [25].

With this preliminaries, in this paper we assume a braneworld model
with induced gravity whose bulk action includes, in addition to the
familiar Einstein term, a Gauss-Bonnet contribution. We relax the
mirror symmetry of the embedding and we derive the field equations
on the brane. In section $2$, we derive the bulk solution, which in
general has the Schwarzschild-anti de Sitter form. Using the
generalized junction conditions, we derive the effective Einstein
equations in the bulk and brane. The absence of the mirror symmetry
and the presence of the GB term lead to a novel and very complicated
Friedmann equation in the brane which will be interpreted in several
interesting subcases. Section $3$ deals with the cosmological
dynamics of this asymmetric braneworld setup in the absence of the
GB curvature effect. We will show that in the asymmetric case of
this pure induced gravity scenario, the phantom-like behavior can be
realized in the smaller redshifts than the symmetric case. In
section $4$, we consider the general case with the GB term and we
show that it is possible to realize the phantom-like behavior by
justification of some conditions governed on the field equations.
Finally, our summary and conclusions are presented in section $5$.

\section{Equations of motion}
In this section we consider an asymmetric braneworld model with the
Gauss-Bonnet contribution in the bulk and induced gravity term on
the brane . The action of this model is given as follows
\begin{equation}
{\cal{S}}={\cal{S}}_{bulk}+{\cal{S}}_{brane},
\end{equation}
where
\begin{equation}
{\cal{S}}_{bulk}=\sum_{i=1,2}M_{i}^{3}\int_{{\cal{B}}_{i}}d^{5}x\sqrt{-g}\bigg({\cal{R}}_{i}
-2\Lambda_{i}+\alpha_{i}{\cal{L}}_{GB}^{i}\bigg)-2M_{i}^{3}\int_{\partial{\cal{B}}_{i}}d^{4}x\sqrt{-h}
\bigg[K_{i}+2\alpha_{i}\Big(J-2G^{ab}K_{ab}\Big)\bigg].
\end{equation}
and
\begin{equation}
{\cal{S}}_{brane}=\int_{brane}d^{4}x\sqrt{-h}\Big(m^2R-2\sigma+{\cal{L}}_{M}
\Big)
\end{equation}
Here ${\cal{S}}_{bulk}$ is the action of the bulk,
${\cal{S}}_{brane}$ is the brane action and ${\cal{S}}$ is the total
action. ${\cal{B}}_{i}(i=1,2)$ is corresponding to two bulk spaces
on either sides of the brane with $5$-dimensional gravitational and
cosmological constants $M_{i}^{3}$ and $\Lambda_{i}$ respectively.
${\cal{R}}_{i}$ is the scalar curvature of the bulk metric
$g_{ab}^{i}$. $\alpha_{i}>0$ is the Gauss-Bonnet coupling and
${\cal{L}}_{GB}^{i}$ is the Gauss-Bonnet term in the bulk defined by
$${\cal{L}}_{GB}={\cal{R}}^2-4{\cal{R}}_{ab}{\cal{R}}^{ab}+
{\cal{R}}_{abcd}{\cal{R}}^{abcd}.$$ $h_{ab}$ is the induced metric
on the brane and is given by $h_{ab}=g_{ab}-n_{a}n_{b}$ where
$n^{a}$ is the outward pointing unit normal to
${\partial{\cal{B}}_{i}}$. $K_{i}=K_{ab}^{i}h^{ab}$ is the trace of
the extrinsic curvature of the brane in the bulk. $G^{ab}$ is the
$4$D Einstein tensor on the brane and $J$ is the trace of
\begin{equation}
J_{ab}=\frac{1}{3}\big(2KK_{ac}{K^{c}}_{b}+K_{cd}K^{cd}K_{ab}-2K_{ac}K^{cd}K_{db}-K^2K_{ab}\big).
\end{equation}
$R$ is the induced scalar curvature on the brane and $\sigma$ and
${\cal{L}}_{M}$ are the tension and matter Lagrangian of the brane
respectively.

Variation of the action gives the following field equations
\begin{equation}
{\cal{G}}_{ab}^{i}+\alpha
H_{ab}^{i}+\Lambda_{i}g_{ab}^{i}=\frac{1}{2M_{i}^3}S_{ab}^{i}\delta(y)
\end{equation}
where $H_{ab}$ is the Lovelock tensor defined by
\begin{equation}
H_{ab}={\cal{R}}{\cal{R}}_{ab}-{\cal{R}}_{ac}{{\cal{R}}^{c}}_{b}-2{\cal{R}}^{cd}{\cal{R}}_{abcd}
+{{\cal{R}}_{a}}^{cde}{\cal{R}}_{bcde}-\frac{1}{4}g_{ab}{\cal{L}}_{GB}
\end{equation}
and $S_{ab}$ is the contribution from the brane located at $y=0$ and
has the following form [26]
\begin{equation}
S_{ab}=T_{ab}^{brane}-\sigma h_{ab}-m^2G_{ab}
\end{equation}
The stress energy in the brane is
$T_{ab}^{brane}=\frac{2\delta{\cal{L}}_{M}}{\delta
h^{ab}}-h_{ab}{\cal{L}}_{M}.$ Using the Gauss-Codazzi equation, the
energy momentum of the matter on the brane is conserved [27] and
therefore $\nabla^{a}T_{ab}=0$, where $\nabla^{a}$ is the covariant
derivative on the brane associated with the induced metric $h_{ab}$.
Note that this relation is a consequence of the absence of matter in
the bulk.

\subsection{Bulk Solution}
At this stage, let us to relax the index $i$ since the following
analysis will apply to both subspaces of the bulk manifold. We will
save this index when it is necessary. For a homogeneous and
isotropic brane, the bulk metric can be written, using the
generalized Birkhoff's theorem, in the following form [28]
\begin{equation}
ds^2=-f(a)dt^2+\frac{da^2}{f(a)}+a^2\gamma_{ij}dx^{i}dx{j}
\end{equation}
Here, $\gamma_{ij}$ is a three dimensional metric of a space with
constant curvature $k=-1,0,1$. With this metric, the $(i,i)$ and
$(t,t)$ or $(a,a)$ components of the field equations (5) are given
as [29,30]
\begin{equation} \Big(a^2+4\alpha
k-4\alpha\Big)\frac{d^2f}{da^2}-4\alpha\Big(\frac{df}{da}\Big)^2+2\Big(f+\Lambda
a^2-k\Big)=0
\end{equation}
and
\begin{equation}
\alpha\Big(3af-2k\Big)\frac{df}{da}-3fa^2-\Lambda a^4+3ka^2=0\,,
\end{equation}
respectively. Note that the latter equation acts as a constraint
equation. The solution of these equation is given by [29]
\begin{equation}
f(a)=k+\frac{a^2}{4\alpha}\Big(1\mp\sqrt{1+\frac{4}{3}\alpha\Lambda+8\alpha\frac{\mu}{a^4}}\Big)
\end{equation}
where $\mu\geq0$ is an arbitrary constant related to the black hole
mass by the relation $M_{BH}=3M^3V\mu$, where $V$ is the volume of
the $3$D space [28]. The solution for $f(a)$ has two branches due to
two signs in front of the square root in equation (11). For the
negative branch, the solution has the general relativistic limit as
$\alpha\rightarrow0$, but for positive branch there is no general
relativistic limit ( see [29] and references therein ). For
$\mu\neq0$, the metric (8) faces two classes of singularities.
Firstly, an essential singularity at $a=0$. For the negative branch,
this singularity is covered by an event horizon if $k\leq0$ or $k=1$
and $\mu\geq2\alpha$. For such cases, the event horizon ($a=a_{h}$)
is
$$a_{h}^2=\frac{3k}{\Lambda}+\sqrt{\frac{9k^2+12k^2\alpha\Lambda-6\mu\Lambda}{\Lambda^2}}.$$
However, this is not the case for the positive branch. So, to
discard this naked singularity we must cut the spacetime off at some
small values of $a$. This can be done by introducing a second brane
at $a\sim M_{cut}^{-1}$ [29]. The second class of singularities, the
so called \emph{branch singularities}, occur when
$a=a_{b}:=\big[-\alpha\mu/\big(1+\frac{4\alpha\Lambda}{3}\big)\big]^{1/4}>0$,
in which the term inside the square root in equation (11) vanishes.
In order to avoid this singularity, one requires
$1+\frac{4\alpha\Lambda}{3}\geq0$ and $\alpha\mu\geq0$. From this
relation, since $\mu\geq0$ we must have $\alpha\geq0$ which is
consistent with string theory considerations.

\subsection{Friedmann equation on the Brane}
We consider the location of the brane as $t=t(\tau)$ and
$a=a(\tau)$, which is parameterized by the proper time $\tau$ of the
brane. Then the induced metric on the brane is given by
\begin{equation}
ds^2=-d\tau^2+a_{i}(\tau)^2\gamma_{ij}dx^{i}dx{j}
\end{equation}
In this equation, $\tau$ and $a_{i}(\tau)$ are corresponding to the
cosmic time and the scale factor of the Friedmann-Robertson-Walker
universe respectively. Note that, since the brane coinsides with
both boundaries, the metric on the brane is only well defined when
$a_{1}(\tau)=a_{2}(\tau)=a(\tau)$. The Hubble parameter on the brane
is defined by $H=\frac{\dot{a}}{a}$\,\,. The tangent vector at a
point on the brane can be written as follows
\begin{equation}
u=\dot{t}_{i}\frac{\partial}{\partial
t_{i}}+\dot{a}\frac{\partial}{\partial a}
\end{equation}
where $i=1,2$ corresponds to two sides of the brane and a dot marks
a derivative with respect to the proper time $\tau$. The normal
vector to the brane is
$$n_{a}^{(i)}=\theta_{i}\Big(-\dot{a}_{i}(\tau),\dot{t}_{i}(\tau),0\Big)$$
where $\theta_{i}=\pm1$. For $\theta_{i}=1$, ${\cal{B}}_{i}$
corresponds to $0\leq a<a_{i}(\tau)$ whereas for $\theta_{i}=-1$,
${\cal{B}}_{i}$ is corresponding to $a_{i}(\tau)<a<\infty$. With
this definition, the conditions  $n_{\mu}u^{\mu}=0$ and
$n_{\mu}n^{\mu}=1$ are satisfied. Normalization of $n^{a}$ imposes a
constraint equation so that
\begin{equation}
-f_{i}(a)\dot{t}^{2}+\frac{\dot{a}^{2}}{f_{i}(a)}=-1.
\end{equation}
The dynamics of the brane is given by the generalized junction
conditions for a Gauss-Bonnet braneworld gravity [31]
\begin{equation}
[K_{ab}]-h_{ab}[K]+2\bigg(3[\alpha J_{ab}]-h_{ab}[\alpha
J]-2P_{abcd}[\alpha K^{ab}]\bigg)=-S_{ab}
\end{equation}
where
\begin{equation}
P_{abcd}=R_{abcd}+2h_{a[d}R_{c]b}+2h_{b[c}R_{d]a}+Rh_{a[c}h_{d]b}
\end{equation}
is the divergence-free part of the Riemann tensor of the brane which
can be constructed from the induced metric on the brane, $h_{ab}$.
Note that by definition, $[X]\equiv X_{2}R_{2}-X_{1}R_{1}$\,\, and
$S_{ab}$ is defined as equation (7). We take the brane matter to be
a perfect fluid, so $T_{ab}=(\rho+p)u_{a}u_{b}+ph_{ab}$, where
$\rho$ and $p$ are the energy density and pressure of the perfect
fluid respectively. The extrinsic curvature has the following
non-vanishing components
\begin{equation}
(K_{i})_{ab}u^{a}u^{b}=\theta_{i}(h_{i}\dot{t}_{i})^{-1}
\Big(\ddot{a}+\frac{\dot{h}_{i}}{2}\Big)
\end{equation}
and
\begin{equation}
(K_{i})_{r}^{r}=(K_{i})_{\theta}^{\theta}=(K_{i})_{\varphi}^{\varphi}=-\theta_{i}\frac{h_{i}\dot{t}_{i}}{a}.
\end{equation}
Now, the $(\tau,\tau)$ component of the equation (15) can be recast
as follows
\begin{equation}
\sum_{i=1,2}\theta_{i}M_{i}^3\frac{f_{i}\dot{t}_{i}}{a}\bigg[1-\frac{4}{3}
\alpha(\frac{f_{i}\dot{t}_{i}}{a})^2+4\alpha(H^2+\frac{k}{a^2})\bigg]
=\frac{\rho+\sigma}{3}-m^2(H^2+\frac{k}{a^2})
\end{equation}
Using the constraint equation (14), this equation can be rewritten
in the following form
\begin{equation}
H^2+\frac{k}{a^2}=\frac{\rho+\sigma}{3m^2}+\frac{1}{m^2}\sum_{i=1,2}\theta_{i}M_{i}^3\sqrt{\frac{f_{i}(a)}{a^2}+H^2}
\bigg[1+\frac{8\alpha_{i}}{3}H^2+\frac{4\alpha_{i}}{3}\bigg(\frac{3k-f_{i}(a)}{a^2}\bigg)\bigg].
\end{equation}
This is a very complicated equation for cosmological dynamics on the
brane. In which follows, we consider some especial and simpler cases
of this equation to study the possible realization of the
phantom-like behavior on the brane.

\subsection{A pure induced gravity brane ($\alpha=0$) without $Z_{2}$ symmetry }
In this section we consider a pure induced gravity scenario in the
absence of the mirror symmetry and the GB contribution ($\alpha=0$).
Then we investigate the phantom-like behavior of this model. In this
case $f(a)$ has the following form
\begin{equation}
f(a)=k\pm\Big(\frac{\Lambda}{6}a^2+\frac{\mu}{a^2}\Big).
\end{equation}
Only the negative sign of  this relation is well-defined, so we take
this branch. Using equations (14) and (21), the field equation (20)
can be recast in the following form
\begin{equation}
H^2+\frac{k}{a^2}=\frac{\rho+\sigma}{3m^2}+\frac{1}{m^2}\sum_{i=1,2}\theta_{i}M^3_{i}
\sqrt{H^2+\frac{k}{a^2}-\frac{\Lambda_i}{6}-\frac{\mu_i}{a^4}}\,.
\end{equation}
In which follows, we ignore the dark radiation term $(\mu_i/a^4)$,
because we investigate the cosmological implications of the model at
the late epoches and this term decays at late times very fast. This
term is important when one treats perturbations on the brane. To
investigate the cosmological implications, we consider the Friedmann
equation (22) in some especial cases. Note that the general case has
been studied in Ref [13].

\subsubsection{The Case with  $\Lambda_{1}=\Lambda_{2}=\Lambda_{b}$ , \quad$M_{1}\neq M_{2}$}
In this case the Friedmann equation (22) has a general solution as
follows
\begin{equation}
H^2+\frac{k}{a^2}=\frac{\rho+\sigma}{3m^2}+2\Big(\frac{1}{l_1}+\frac{1}{\epsilon
l_2}\Big)^2\bigg[ 1\pm\sqrt{1+\Big(\frac{l_1l_2}{l_1+\epsilon
l_2}\Big)^2\Big(\frac{\rho+\sigma}{3m^2}-\frac{\Lambda_b}{6}\Big)}\bigg]
\end{equation}
where we have defined the length scale $l_i=\frac{2m^2}{M_i^3}$ and
$\epsilon=\theta_1\theta_2$. Also we have assumed $M_1>M_2>0$. Note
that for $\epsilon=1$, these solutions are a generalization of
the\emph{ brane $1$} and \emph{brane $2$} branches studied in
reference [32] for asymmetric brane. However in our scenario we have
two extra branches due to asymmetric embedding of the brane. In this
regard we can define an effective cosmological constant on the brane
as follows
\begin{equation}
H^2=\frac{\rho}{3m^2}+\frac{\Lambda_{eff}}{3}
\end{equation}
where we have assumed a flat FRW brane and
\begin{equation}
\frac{\Lambda_{eff}}{3}=\frac{\sigma}{3m^2}+2\Big(\frac{1}{l_1}+\frac{1}{\epsilon
l_2}\Big)^2\bigg[1\pm\sqrt{1+\Big(\frac{l_1l_2}{l_1+\epsilon
l_2}\Big)^2\Big(\frac{\rho+\sigma}{3m^2}-\frac{\Lambda_b}{6}\Big)}\bigg].
\end{equation}
This equation indicates that a late time behavior can be deduced in
all branches of the scenario. Choosing the lower sign in the right
hand side of the equation (25) leads to a very interesting result.
Indeed, in this case the effective cosmological constant can be
decomposed into two distinct parts: the first term on the rhs of
(25) can be considered as a cosmological constant on the brane and
the second term has an screening effect. The screening effect is due
to the induced gravity term on the brane that leads to increasing of
the effective cosmological constant with cosmic time. So, there are
two branches of the scenario that give us an opportunity to
investigate a phantom-like behavior on brane. With phantom-like
behavior, we mean growth of the effective dark energy density with
cosmic time and the effective equation of state parameter that must
be less than $-1$ and at the same time the Hubble rate $\dot{H}$
must be negative to avoid a Big Rip type singularity in the future.
To realize a phantom-like behavior, we rewrite the Friedmann
equation (23) for the lower sign in a dimensionless form
\begin{equation}
E^2(z)=\frac{H^2}{H^2_0}=\Omega_m(1+z)^3+\Omega_{\sigma}+2A-
2\sqrt{A}\sqrt{\Omega_m(1+z)^3+\Omega_{\sigma}+\Omega_{\Lambda_b}+A}
\end{equation}
where we have defined $A=\Omega_{l_1}+\Omega_{l_2}
+2\epsilon\sqrt{\Omega_{l_1}\Omega_{l_2}}$ and the cosmological
parameters are defined as:
$\Omega_{m}=\frac{\rho_{0m}}{3m^{2}H_{0}^{2}}$\, ,\,
$\Omega_{\sigma}=\frac{\sigma}{3m^{2}H_{0}^{2}}$\, ,\,
$\Omega_{l_i}=\frac{1}{l_i^{2}H_{0}^{2}}$\,  ,\,
$\Omega_{\Lambda_b}=\frac{-\Lambda_b}{6H_{0}^{2}}$\,\, and $\rho_0$
\, and \, $H_0$ \, are the present values of matter energy density
and Hubble parameter respectively. Note that a constraint equation
is imposed on the dimensionless equation (26) at redshift $z=0$, so
that
\begin{equation}
1=\Omega_m+\Omega_{\sigma}+2A-
2\sqrt{A}\sqrt{\Omega_m+\Omega_{\sigma}+\Omega_{\Lambda_b}+A}.
\end{equation}
To investigate the phantom-like behavior, we rewrite the standard
Friedman equation as follows
\begin{equation}
H^{2}=\frac{1}{3m^{2}}\Big(\rho+\rho^{(DE)}_{eff}\Big)
\end{equation}
where $\rho$ is the energy density of the standard matter and
$\rho^{(DE)}_{eff}$  is energy density corresponding to dark energy.
Taking the negative sign of equation (23) and comparing it with
equation (28) leads to the following relation for
$\rho^{(DE)}_{eff}$
\begin{equation}
\rho^{(DE)}_{eff}=\sigma+3m^2\Big(\frac{1}{l_1}+\frac{1}{\epsilon
l_2}\Big)^2\bigg[1-\sqrt{1+\Big(\frac{l_1l_2}{l_1+\epsilon
l_2}\Big)^2\Big(\frac{\rho+\sigma}{3m^2}-\frac{\Lambda_b}{6}\Big)}\bigg]\,.
\end{equation}
Now, the Hubble rate is
\begin{equation}
\dot{H}=\frac{-\rho}{2m^2}\bigg(1-\frac{1}{\sqrt{1+\Big(\frac{l_1l_2}{l_1+\epsilon
l_2}\Big)^2\Big[\frac{\rho+\sigma}{3m^2}-\frac{\Lambda_b}{6}\Big]}}\bigg)\,.
\end{equation}
This relation shows that $\dot{H}<0$. By computing time derivative
of equation (28) and using the fact that $\dot{H}<0$, we find that
$\dot{\rho}_{eff}>0$. Note that $\rho^{(DE)}_{eff}$ satisfies the
energy conservation equation
\begin{equation}
\dot{\rho}_{eff}+3H(1+\omega_{eff})\rho_{eff}=0\,,
\end{equation}
where the effective equation of state parameter is defined as
$\omega_{eff}=\frac{P_{eff}}{\rho_{eff}}$ and $P_{eff}$ is the
effective pressure of the dark energy component. In this respect,
the effective equation of state parameter of dark energy can be
expressed as follows
\begin{equation}
1+\omega_{eff}=\frac{-\Omega_{m}(1+z)^{3}}{\bigg(E^{2}-
\Omega_{m}(1+z)^{3}\bigg)\sqrt{1+\frac{1}{A}\Big(\Omega_{m}(1+z)^{3}+
\Omega_{\sigma}+\Omega_{\Lambda_b}\Big)}}
\end{equation}
In figure $1$ we have plotted the $\omega_{eff}$ versus the
redshift. This figure shows that, for both symmetric
($\theta_1=\theta_2=-1$) and asymmetric ($\theta_1=-\theta_2=1$)
case, the effective equation of state parameter stays in the phantom
region. However, there is no smooth crossing of the phantom divide
line, $\omega_{eff}=-1$, in this setup. Indeed, to have a smooth
crossing of the phantom divide line, a canonical scalar field should
be added to the brane action ( see the first reference of [23]). It
is important to note that the asymmetric effect leads to a breakdown
of the effective phantom-like picture in smaller redshifts relative
to the phantom-like behavior on the symmetric brane.
\begin{figure}[htp]
\begin{center}\includegraphics{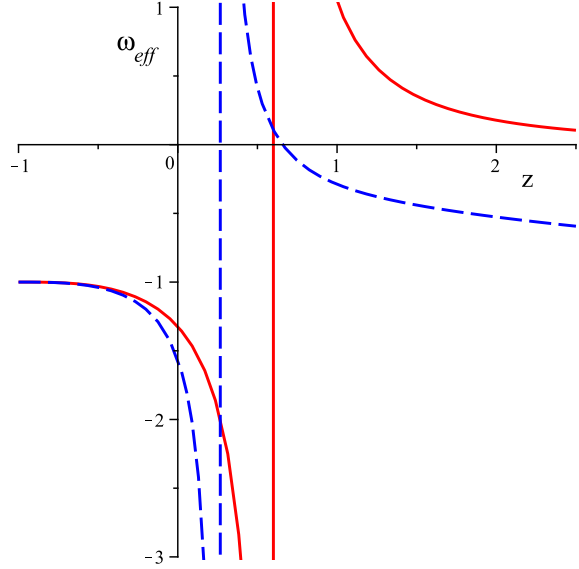} \vspace{6.5cm}
\end{center}
 \caption{\small {Variation of the effective equation of state parameter versus the
redshift for the symmetric case ( solid line )and asymmetric case (
dashed line). For both symmetric ($\theta1=\theta_2=-1$) and
asymmetric ($\theta1=-\theta_2=1$) case, it is possible for
effective equation of state parameter to lie in the phantom region.
}}
\end{figure}
The deceleration parameter is given by
$q=-(1+\frac{\dot{E}}{H_{0}E^{2}})$\, where
\begin{equation}
\frac{\dot{E}}{H_{0}}=-\frac{3}{2}\Omega_{m}(1+z)^{3}
\Bigg(1-\sqrt{\frac{A}{A+\Omega_{m}(1+z)^{3}+
\Omega_{\lambda}+\Omega_{\Lambda_b}}}\Bigg)< 0
\end{equation}
This relation implies that the deceleration parameter never can be
less than $-1$. Consequently, there is no super-acceleration and big
rip singularity in this model.  Figure $2$ shows the variation of
the deceleration parameter versus the redshift for both asymmetric
and symmetric branches. As it is obvious, the deceleration parameter
reduces by redshift towards the recent epoch. Here the deference is
in such a way that for the asymmetric case the deceleration
parameter vanishes in the smaller redshift. This means that the
accelerating phase in the symmetric case lasts more than the
asymmetric case. Since the self-accelerating behavior of the
universe for the asymmetric case is started in the smaller redshist
relative to the symmetric case, it seems that the asymmetric case is
more suitable to realize the phantom-like behavior. This is
reasonable since the model parameter space is wider in the
asymmetric case.
\begin{figure}[htp]
\begin{center}\includegraphics{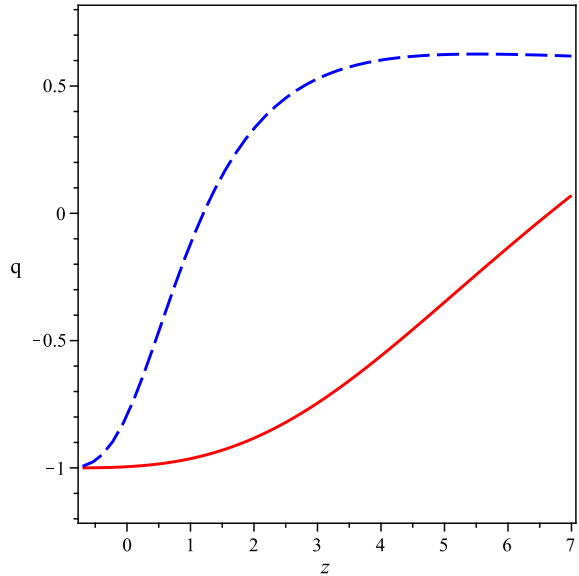} \vspace{7cm}
\end{center}
 \caption{\small {Variation of the deceleration
parameter versus the redshift for the symmetric case ( solid line )
and asymmetric case ( dashed line ). For both cases the deceleration
parameter reduces with redshift towards the recent epoch.}}
\end{figure}

\subsubsection{The Case with $\alpha=0$ and $m=0$}
The case without induced gravity ($m=0$) has been studied thoroughly
in the literature ( see for instance [15] and also [12]). Here we
give a brief review of this case for completeness and in addition
comment on the possible variation of the fundamental scales. It has
been shown that for $\theta_1\theta_2=1$ this model is corresponding
to the generalized Randall-Sundrum (RS) model (with
$\theta_1=\theta_2=1$) and inverse RS model
($\theta_1=\theta_2=-1$). However, for $\theta_1\theta_2=1$ this
case cannot lead to a de-sitter phase and therefore it cannot be
accounted for the late-time acceleration of the universe. For
$\theta_1\theta_2=-1$, the Friedmann equation (22) with $m=0$ has
the following form [12]
$$
H^2+\frac{k}{a^2}=\frac{M_1^6\Lambda_1-M_2^6\Lambda_2}{6(M_1^6-M_2^6)}+
\frac{M_1^6+M_{2}^6}{9(M_1^6-M_2^6)}\Big(\rho+\sigma\Big)^2$$
\begin{equation}
\pm\frac{2M_1^3M_2^3}{9(M_1^6-M_2^6)}\Big(\rho+\sigma\Big)\bigg[\Big(\rho+\sigma\Big)^2+
\frac{3}{2}(\Lambda_1-\Lambda_2)(M_1^6-M_2^6)\bigg]^{\frac{1}{2}}.
\end{equation}
In a certain range of parameters, this model is cosmologically
equivalent to an induced gravity scenario in the presence of mirror
symmetry. When $$ \rho+\sigma\ll(\rho+\sigma)_{max}=
\frac{2M_1^3M_2^3}{(M_1^6+M_2^6)}\sqrt{(M_1^6-M_2^6)(\Lambda_1-\Lambda_2)}$$
the Friedmann equation approximates to [15,12]
\begin{equation}
H^2+\frac{k}{a^2}\approx\frac{\rho+\sigma}{3m_b^2}+\frac{\Lambda_{eff}}{6}
\end{equation}
where $$m_b^2=\frac{(M_1^6-M_2^6)^2}{M_1^3M_2^3}\sqrt{\frac{3}
{2(M_1^6-M_2^6)(\Lambda_1-\Lambda_2)}}$$ and
$$\Lambda_{eff}=\frac{M_1^6\Lambda_1-M_2^6\Lambda_2}{(M_1^6-M_2^6)}.$$
Since the effective cosmological constant in this case cannot evolve
with time, it is impossible to realize a phantom-like behavior in
this situation. However, if we consider some sort of evolving matter
fields in the bulk ( such as canonical scalar fields), that can be
different in either sides of the brane, it is possible essentially
to realize a phantom-like behavior on the brane in this case. Also
one can realize the phantom-like behavior in this case even by
incorporation of the possibility to have varying fundamental
constants in either sides of the brane. The importance of the
possible realization of the phantom-like effect in these manner lies
in the fact that this phantom-like behavior will occur in the
absence of the induced gravity. If this is actually the case, it
will be an important progress in this field. These conjectures are
under investigation and will be reported in our forthcoming works.

\subsection{The Case with $\alpha\neq0$}
\subsubsection{A stringy model with induced gravity}
In a stringy model with induced gravity, there is no bare
cosmological constant in the bulk action ($\Lambda_i=0$), there is
no tension in the brane action ($\sigma=0$) and the Gauss-Bonnet
parameter is always positive ($\alpha_i>0$). It is important to note
that in the presence of the mirror symmetry, this case has been
dubbed as \emph{GBIG scenario} and its cosmological dynamics and
effective phantom-like behavior have been studied extensively in the
literature ( see [19] and [33]).

For the positive sign in equation (11), the bulk metric contains
\begin{equation}
f(a)=k+\frac{a^2}{2\alpha}.
\end{equation}
Note that this branch is not well-defined at $\alpha=0$. For the
negative sign of equation (11), the only allowed value of $k$ is
$k=1$, therefore $f(a)=1$. Since we are interested in a flat FRW
universe, we take only the positive branch in our analysis. In this
case, the Friedmann equation is given in the following form
\begin{equation}
H^2=\frac{\rho+\sigma}{3m^2}+\frac{2}{3}\sum_{i=1,2}\frac{\theta_{i}}{l_{i}}
\sqrt{\frac{1}{2\alpha_i}+H^2}\bigg[1+8\alpha_iH^2\bigg]
\end{equation}
where $l_{i}$ has been introduced in the previous section. One of
the simplest cosmological models that can exhibit the late-time
acceleration of the universe is a de Sitter space time. So, it is
worth to seek for such a solution in our model. To do this end, we
set $\rho=0$ which is an appropriate choice for late-times. Then the
Friedmann equation (37) is simplified to
\begin{equation}
\lim_{z\rightarrow-1}
H^2(z)=H^2_{ds}=\frac{2}{3}\sum_{i=1,2}\frac{\theta_{i}}{l_{i}}
\sqrt{\frac{1}{2\alpha_i}+H^2}\bigg[1+8\alpha_iH^2\bigg].
\end{equation}
This equation implies that this model can evolve to a de Sitter
phase at late-times for two branches with $\theta_1=\theta_2=1$ and
$\theta_1\theta_2=-1$ with $M_1>M_2$ for the latter case. Note that
for $m=0$, as has been shown by Padilla [15], only the asymmetric
branch can exhibite a cosmic acceleration. Here, the effect of the
induced gravity term leads to a de Sitter phase for the symmetric
branch with $\theta_1=\theta_2=1$. To investigate the phantom
mimicry of this model, similar to the previous section we define an
effective cosmological constant on the brane due to extra
dimensional effects as follows
\begin{equation}
\frac{\Lambda_{eff}}{3}=\frac{\sigma}{3m^2}+\frac{2}{3}\sum_{i=1,2}\frac{\theta_{i}}{l_{i}}
\sqrt{\frac{1}{2\alpha_i}+H^2}\bigg[1+8\alpha_iH^2\bigg]
\end{equation}
For a phantom accelerating phase,\,  $\omega(z)<-1$\,, that by
definition [32]
\begin{equation}
\omega_{z}=\frac{2q(z)-1}{3[1-\Omega_{m}(z)]}\, ,\,\quad
q(z)=\frac{d \log H(z)}{d \log (1+z)}-1\, ,\, \quad
\Omega_{m}(z)=\frac{\Omega_{m}(1+z)^3}{E^2(z)}.
\end{equation}
This condition is equivalent to\,\,
$\Omega_{m}(z)>\frac{2}{3}\frac{d \log H(z)}{d \log (1+z)}$\quad and
$\dot{\Lambda}_{eff}>0$\,,\, where a dot marks a differentiation
with respect to the cosmic time.  The expansion rate $\dot{H}$ is
given by
\begin{equation}
\dot{H}=-\frac{\rho}{2m^2}\Bigg[1-\frac{1}{3}\sum_{i=1,2}\frac{\theta_{i}}{l_{i}}
\bigg(\frac{9+24\alpha_iH^2}{\sqrt{\frac{1}{2\alpha_i}+H^2}}
\bigg)\Bigg]^{-1}.
\end{equation}
For
\begin{equation}
\sum_{i=1,2}\frac{\theta_{i}}{l_{i}}
\bigg(\frac{9+24\alpha_iH^2}{\sqrt{\frac{1}{2\alpha_i}+H^2}}\bigg)<0,
\end{equation}
we find $\dot{H}<0$. Since the deceleration parameter is related to
$\dot{H}$ via relation $q=-\Big(1+\frac{\dot{H}}{H^2}\Big)$\,, with
$\dot{H}<0$ we find $q>-1$. As a result, there is no
superacceleration in this case. Since $\dot{H}<0$, the effective
cosmological constant increases with time
\begin{equation}
\dot{\Lambda}_{eff}=\frac{2}{3}H\dot{H}\sum_{i=1,2}\frac{\theta_{i}}{l_{i}}
\bigg(\frac{9+24\alpha_iH^2}{\sqrt{\frac{1}{2\alpha_i}+H^2}}\bigg)>0
\end{equation}
On the other hand, $\omega_{eff}$ can be obtained by virtue of the
continuity equation
\begin{equation}
\omega_{eff}=-\Big(1+\frac{\dot{\Lambda}_{eff}}{\Lambda_{eff}}\Big)<-1.
\end{equation}
Note that the phantom-like behavior can be realized in this case for
$\theta_1=\theta_2=-1$ and also one of the mixed branches
(\emph{i.e.} $\theta_1\theta_2=-1$).

\subsubsection{The general case with $\Lambda_i\neq0$ and $\sigma\neq0$}
Now, we consider a more general case. For a spatially flat universe
without dark radiation term, the solutions for the bulk metric are
given by
\begin{equation}
f_i(a)=\frac{a^2}{4\alpha_i}\bigg(1\mp\sqrt{1+\frac{4}{3}\alpha_i\Lambda_i}\bigg)
\end{equation}
Then, the Friedmann equation (20) can be rewritten in the following
form
\begin{equation}
H^2=\frac{\rho+\sigma}{3m^2}+\frac{2}{3}\sum_{i=1,2}\frac{\theta_{i}}{l_{i}\sqrt{\alpha_{i}}}
\sqrt{1+4\alpha_{i}H^2\mp\sqrt{1+\frac{4}{3}\alpha_i\Lambda_i}}
\Bigg(1+4\alpha_{i}H^2\pm\frac{1}{2}\sqrt{1+\frac{4}{3}\alpha_i\Lambda_i}\Bigg)
\end{equation}
There are some constraints imposed on this equation. One of these
constraints is $\Lambda_i<-\frac{3}{4}\alpha_i$. Note that, since
the cosmological constant in the bulk is negative, this constraint
is satisfied naturally for positive $\alpha_{i}$'s. On the other
hand, by choosing the negative sign in relation (45), the square
root of equation (46) should be positive. Therefore, we find
\begin{equation}
1+4\alpha_{i}H^2-\sqrt{1+\frac{4}{3}\alpha_i\Lambda_i}>0
\end{equation}
It can be checked easily that this condition is satisfied too. Now,
by adopting a strategy much similar to the procedure that has led us
to equation (25), the effective cosmological constant in this setup
can be defined as follows
\begin{equation}
\frac{\Lambda_{eff}}{3}=\frac{\sigma}{3m^2}+\frac{2}{3}\sum_{i=1,2}\frac{\theta_{i}}{l_{i}\sqrt{\alpha_{i}}}
\sqrt{1+4\alpha_{i}H^2\mp\sqrt{1+\frac{4}{3}\alpha_i\Lambda_i}}
\Bigg(1+4\alpha_{i}H^2\pm\frac{1}{2}\sqrt{1+\frac{4}{3}\alpha_i\Lambda_i}\Bigg).
\end{equation}
In an analogous manner to the previous subsections, the conditions
for realization of the phantom-like behavior are given as follows
\begin{equation}
\dot{H}=-\frac{\rho}{2m^2}\Bigg[1-\frac{8}{3}\sum_{i=1,2}\frac{\theta_{i}\sqrt{\alpha_i}}{l_{i}}
\bigg(\frac{1+4\alpha_{i}H^2\mp\frac{1}{4}\sqrt{1+\frac{4}{3}\alpha_i\Lambda_i}}
{\sqrt{1+4\alpha_{i}H^2\mp\sqrt{1+\frac{4}{3}\alpha_i\Lambda_i}}}\bigg)\Bigg]^{-1}<0
\end{equation}
and
\begin{equation}
\dot{\Lambda}_{eff}=\frac{16}{3}H\dot{H}\sum_{i=1,2}\frac{\theta_{i}\sqrt{\alpha_i}}{l_{i}}
\bigg(\frac{1+4\alpha_{i}H^2\mp\frac{1}{4}\sqrt{1+\frac{4}{3}\alpha_i\Lambda_i}}
{\sqrt{1+4\alpha_{i}H^2\mp\sqrt{1+\frac{4}{3}\alpha_i\Lambda_i}}}
\bigg)>0
\end{equation}
if the condition
\begin{equation}
\sum_{i=1,2}\frac{\theta_{i}\sqrt{\alpha_i}}{l_{i}}
\bigg(\frac{1+4\alpha_{i}H^2\mp\frac{1}{4}\sqrt{1+\frac{4}{3}\alpha_i\Lambda_i}}
{\sqrt{1+4\alpha_{i}H^2\mp\sqrt{1+\frac{4}{3}\alpha_i\Lambda_i}}}\bigg)<0
\end{equation}
is satisfied. As an important result, in this general case the
phantom-like behavior can be realized with both signs of equation
(46) in the symmetric case ( with $\theta_1=\theta_2=-1$). This can
be happened also in one of the asymmetric branches ( with
$\theta_1\theta_2=-1$) without introducing a phantom field in the
brane or bulk action.

\section{Summary and Conclusion}
In this paper we have assumed a braneworld model with induced
gravity whose bulk action includes, in addition to the familiar
Einstein term, a Gauss-Bonnet contribution. We have relaxed the
mirror symmetry of the embedding, so the gravitational and
cosmological constants and even the Gauss-Bonnet parameter can be
different in either sides of the brane. We have derived the bulk
solutions, which in general have the schwarzschild anti de Sitter
form. Using the generalized junction condition, we have derived the
effective Einstein equation in the bulk and brane. The absence of
the mirror symmetry and the presence of the curvature effects due to
GB correction term leads to a complicated Friedmann equation for
cosmological dynamics on the brane. To find some intuition on the
cosmological dynamics, we have considered the scenario in some
especial cases. Firstly, we consider the case of a pure induced
gravity scenario ( in the absence of the Gauss-Bonnet contribution
in the bulk). We have shown that in this case, an effective
cosmological constant can be defined on the brane which is screened,
but the screening term reduces in time for two branches of the
scenario so that the value of the effective cosmological constant
increases with cosmic time. Thus the phantom like behavior can be
realized in two branches of the scenario, which one of them survives
in the case of mirror symmetry ( this branch is corresponding to the
\emph{Brane1} solution of Ref.[32]) and the other originates from a
pure asymmetric effects (with $\theta_1\theta_2=-1)$. We have shown
that, in the asymmetric case of this pure induced gravity scenario,
the phantom-like behavior can be realized in the smaller redshifts
than the symmetric case. Secondly, we have considered also the
general case with the GB curvature correction and we have shown that
it is possible to realize the phantom-like behavior in this case by
justifying some conditions on the field equations of the scenario.\\

{\bf Acknowledgment}\\
This work has been supported partially by Research Institute for
Astronomy and Astrophysics of Maragha, IRAN.

\end{document}